\documentclass[aps,prd,superscriptaddress,nofootinbib]{revtex4} % for double-spaced preprint
\usepackage[dvips]{graphicx}
\usepackage{bm}        % for math
\usepackage{amssymb}   % for math
\usepackage[utf8]{inputenc}
\usepackage{mathrsfs}
\usepackage{amsmath}
\usepackage{amssymb}
\usepackage{hyperref}
\usepackage{graphicx,subfigure}
\usepackage{color}

\begin{document}

\newtheorem{thm}{Theorem}[section]
\newtheorem{cor}{Corollary}[section]
\newtheorem{lem}{Lemma}[section]
\newtheorem{prop}{Proposition}[section]

\def\const{\text{const.}}
\def\Rnum{{\mathbb R}}
\def\sgn{{\rm sgn}}

\def\eff{\text{eff.}}
\def\peak{\text{peak}}

\def\sech{{\rm sech}}
\def\arctanh{{\rm arctanh}}

\tolerance=50000
\allowdisplaybreaks[4]

%\title{Scattering of sphalerons with a false vacuum in $\phi^6$ model with a symmetric potential} 
\title{False-vacuum bubbles in sphaleron scattering} 
\author{Manuel A. Martínez Sánchez\footnote{ms.manuel.angel@gmail.com}}
\affiliation{A. I. Alikhanyan National Science Laboratory (Yerevan Physics Institute), 2 Alikhanyan Brothers Street, 0036, Yerevan, Armenia}

\author{Christoph Adam\footnote{christoph.adam@usc.es}}
\affiliation{Departamento de Física de Partículas, Universidad de Santiago de Compostela, E-15782 Santiago de Compostela, Spain}
\affiliation{Instituto
Galego de F\'isica de Altas Enerxias (IGFAE) E-15782 Santiago de Compostela, Spain
}

\author{Danial Saadatmand\footnote{saadatmand.d@gmail.com}}
%\email{Saadatmand.d@gmail.com}
\affiliation{A. I. Alikhanyan National Science Laboratory (Yerevan Physics Institute), 2 Alikhanyan Brothers Street, 0036, Yerevan, Armenia}

\begin{abstract}

We investigate the collision dynamics of two bright sphalerons in a (1+1)-dimensional deformed $\phi^6$ scalar field theory with a symmetric potential possessing false vacua. Two one-parameter realizations of the model, referred to as the barrier and well models, are considered and their static and linear instability properties are first reviewed. We then study head-on collisions of boosted sphalerons over a broad range of initial velocities and deformation parameters. The scattering dynamics exhibit a rich variety of final states, including the production of kink-antikink pairs, long-lived oscillons in true and false vacuum, multiple oscillons propagating in false-vacuum regions, and radiative decay. A particularly remarkable outcome is the emergence of a long-lived  bubble of the false broken vacuum bounded by a kink-antikink pair, which repeatedly collapses and re-expands before eventually decaying into an oscillon. These results demonstrate that deformed  $\phi^6$ theories with false vacua exhibit considerably richer sphaleron dynamics than previously known and provide new insight into the role of unstable localized configurations in nonlinear field theories.
\end{abstract}
 \maketitle

\maketitle

\section{Introduction}
Sphalerons are localized and unstable static solutions in nonlinear field theories, commonly identified in various theoretical frameworks \cite{Manton:1988az,MateosGuilarte:1992kil,Shnir:2009ct,Manton:2019qka,Manton:2023mdr}, including the Standard Model of particle physics \cite{Klinkhamer:1990ik,Kunz:1988sx}. Sphalerons also appear in theories involving solitons, such as monopole–antimonopole pairs \cite{Taubes:1982ie,Taubes:1982ieII,Saurabh:2019rrp}, and within the framework of the Skyrme model \cite{Krusch:2004uf,Isler:1989tt,Shnir:2011zza,Shnir:2013ova}. They are of particular importance in the study of baryon number violation within electroweak theory \cite{Klinkhamer:1984di}, as a stable fixed point of the flow in the large flow time in the Higgs model \cite{Hamada:2020rnp} and in two Higgs doublet theories \cite{Grant:2001at}. 
Sphalerons can influence the dynamics of nonlinear field theories in two generic ways. On the one hand, they can show up as intermediate, transient states in the scattering of solitonic or other particle-like field configurations, where they may have a strong impact on the time evolution and final states. On the other hand, despite their instability they may present particle-type behavior like scattering on their own, at least at time scales which are smaller than their typical life times. The present investigation is dedicated to this second type of phenomena. 

It has been indicated that sphalerons are rather ubiquitous in non-linear field theories \cite{Manton:2019qka} and of high physical importance \cite{Klinkhamer:1984di,Hamada:2020rnp,Grant:2001at}. As is frequently the case for nonlinear phenomena, their analysis is simpler in low-dimensional field theories, to which we shall restrict our investigations. The idea is that such low-dimensional analysis allows one to identify and study some generic features of sphalerons, which can also be expected to occur in higher-dimensional systems.
With this in mind, it has, e.g., been shown in low-dimensional models that sphalerons can serve as a source of internal degrees of freedom, which are responsible for the fractal structure observed in the final state formation pattern in kink-antikink collisions \cite{Adam:2021fet}.
Further, a sphaleron without a shape mode or a discrete positive-energy vibrational mode can decay into an oscillon, which exhibits a characteristic double-oscillation structure \cite{Oles:2023ujf}. The dynamics of a semi-BPS sphaleron interacting with a background defect (impurity), where there is no static force between the sphaleron and the impurity, has been explored in \cite{Alonso-Izquierdo:2023shi}.
In addition, the decay of sphalerons has recently attracted considerable attention due to the rich nonlinear dynamics that emerge from their intrinsic instability. The derivative of a sphaleron field yields a zero-mode eigenfunction with a node, but the Sturm oscillation theorem requires the ground-state eigenfunction to be nodeless. Hence, the true ground state must have a negative eigenvalue, rendering the sphaleron linearly unstable under that perturbation \cite{Manton:2023afn,Anco:2025egc}. Two distinct evolution channels have been identified for the sphaleron \cite{Manton:2023afn}. Under a negative perturbation, it collapses into a long-lived oscillon, a localized oscillatory solution \cite{Gleiser:1993pt,Gleiser:2009ys,Navarro-Obregon:2024ieb}; under a positive perturbation, its height increases until the peak reaches the true vacuum, after which the width expands, creating an accelerating kink--antikink pair whose separation increases with time and asymptotically approaches the speed of light, with the energy density accumulating near the expanding interfaces, indicating the onset of a gradient blow-up at late times \cite{anco2026longtimebehavioursphaleronsphi4}. 

The collision of two metastable sphalerons (also referred to as lumps) in a false vacuum was investigated in a deformed $\phi^4$ model with a one-parameter non-symmetric potential \cite{Lima:2021jxl}. It was shown that, depending on the value of the deformation parameter, the collision exhibits a rich variety of outcomes, including the production of one or two kink--antikink-like pairs and the formation of multiple oscillons (up to four). More recently, the dynamics of oscillon--oscillon collisions in the same model have also been investigated \cite{campos2026collisiondynamicsfalsevacuumoscillons}. It was found that oscillons can pass through one another, reflect, merge into a larger long-lived oscillon, generate additional oscillons, or evolve into kink--antikink pairs. These results highlight the rich nonlinear dynamics supported by this class of non-symmetric scalar field theories.

 In this paper, we investigate the interaction between two bright sphalerons of deformed $\phi^6$ models with false vacua, which were introduced in Ref.~\cite{Navarro-Obregon:2024ieb}. The two models, known as the barrier and well models, are described by one-parameter families of symmetric potentials. In the barrier model, the sphaleron is found not to support any bound states, whereas in the well model it supports one or several bound states, depending on the model parameter~\cite{Navarro-Obregon:2024ieb}. We want to emphasize that if there are no impurities or inhomogeneities, then a sphaleron is, in principle, just a traveling wave which does not change shape. However, in numerical simulations, there is always a numerical error which acts to trigger the unstable mode. It is desirable to stress that we consider collisions between them if the instability is relatively slow, so that the sphalerons have "enough time" to collide before being destroyed by the intrinsic instability. 

More concretely, we examine the main characteristics of the scattering products and their dependence on the model parameter and the initial velocity of each sphaleron. In particular, we demonstrate a novel collision outcome in which a long-lived field configuration, referred to as a bubble, becomes temporarily trapped between a separating kink--antikink pair that subsequently propagates toward spatial infinity. This phenomenon occurs only within a specific range of initial collision velocities. To the best of our knowledge, this phenomenon has not previously been reported in sphaleron scattering in real scalar field theories. However, in Ref.~\cite{martinez2025oscillonsbubblesqballdynamics}, the authors observed the formation of short-lived bubbles in Q-ball dynamics in complex scalar theories, where these transient bubble-like configurations can be temporarily stabilized. Their stabilization was attributed to the excitation of massless Goldstone modes, which either exert an outward pressure on the bubble walls or become trapped as bound states within the bubble, thereby delaying its collapse. Although the underlying stabilization mechanism in our model may differ, the appearance of transient bubbles in sphaleron scattering suggests that such localized field configurations may play an important role in the nonlinear dynamics of topological defects.

The structure of the paper is as follows: In Secs.~\ref{sec:II} and~\ref{Sec:LinST} we briefly review the results of Ref.~\cite{Navarro-Obregon:2024ieb} which we need in the following. Concretely, in  Sec.~\ref{sec:II}, we present the models and their main equations, and discuss the possible sphaleron solutions in both models that exist within the range of the model parameters. In Sec.~\ref{Sec:LinST}, we review the linear instability of the existing solutions under small perturbations to the static field and discuss their spectrum. The final states resulting from the scattering outcomes after the collision of two sphalerons in the barrier and well model in a false vacuum are discussed in Sec.~\ref{SphaleronScat}. In Sec.~\ref{SphaleronScatPerturb}, the sphalerons are initially perturbed along their unstable eigenmodes before the collision. Finally, Sec.~\ref{Sec:Con} presents our conclusions and summarizes the key findings of this study.

\section{The model and main equations}\label{sec:II}

We consider the $(1+1)$-dimensional real scalar field theory, whose dynamics are governed by the Lagrangian density
\begin{equation}\label{Lagrangian}
    L = \int_{-\infty}^\infty dx \left( \frac{1}{2} \partial_\mu \phi \partial^\mu \phi - V(\phi,s) \right),
\end{equation}
where the potential $V(\phi,s)$ corresponds to a deformed $\phi^6$ model modified by the parameter $s$. We consider two different deformations, referred to as the barrier and well models, which were introduced in Ref.~\cite{Navarro-Obregon:2024ieb}. These potentials are defined as follows

\begin{equation}\label{eq:potentialB}
    V_B(\phi,s) = \frac{1}{2} \phi^2 \left( \tanh s - \phi^2 \right) \left( \coth s - \phi^2 \right),
\end{equation}
\begin{equation}\label{eq:potentialW}
 V_W(\phi,s) = \frac{(\phi - 1)^2 - \tanh^2 s}{8 \left( 1 - \tanh^2 s \right)} \left( (\phi - 1)^2 - 1 \right)^2. 
\end{equation}
%
% %
% \begin{align}\label{eq:potentials}
%     V_B(\phi,s) &= \frac{1}{2} \phi^2 \left( \tanh s - \phi^2 \right) \left( \coth s - \phi^2 \right),\\
%     V_W(\phi,s) &= \frac{(\phi - 1)^2 - \tanh^2 s}{8 \left( 1 - \tanh^2 s \right)} \left( (\phi - 1)^2 - 1 \right)^2. \label{eq:potentials1}
% \end{align}
% %
%
\begin{figure}
    \centering
    \includegraphics[width=\linewidth]{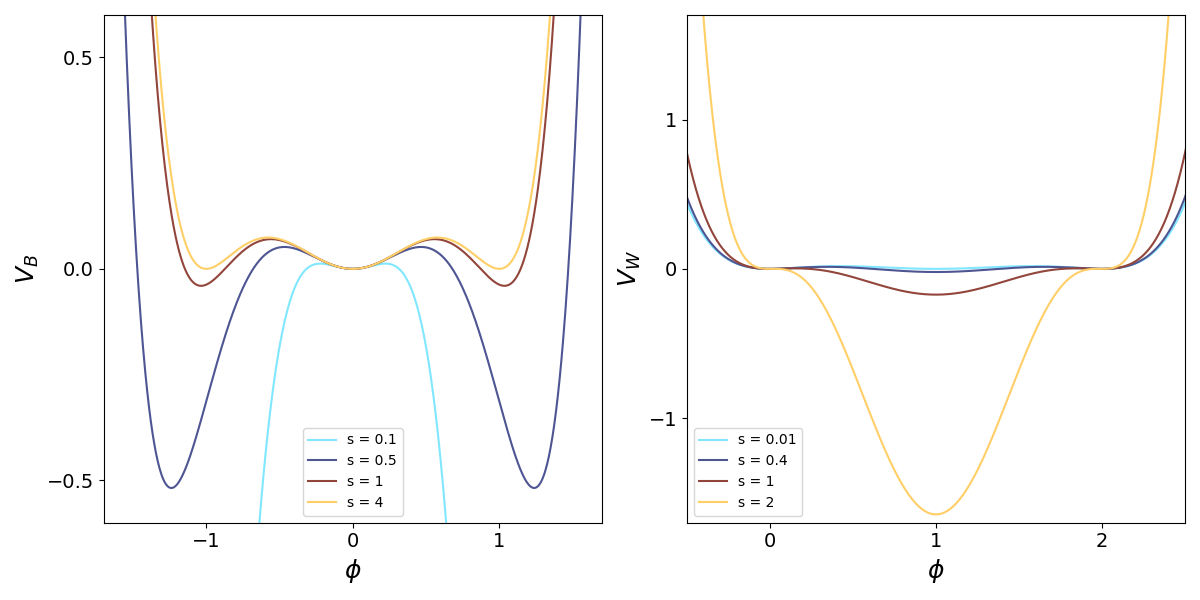}
    \caption{Barrier (left) and well (right) potentials for different values of $s$.}
    \label{fig:Potentials}
\end{figure}

The first potential is called the barrier model and has one false vacuum at $\phi=0$ and two symmetric true vacua. In the limit $s\to\infty$, the model reduces to the standard $\phi^6$ model with three vacua located at $\phi=0$ and $\phi=\pm1$. The second potential has one true vacuum at $\phi=1$ and two false vacua at $\phi=0$ and $\phi=2$. These two types of potentials are plotted in Fig.~\ref{fig:Potentials}.

From the Lagrangian \eqref{Lagrangian} one can obtain the equation of motion for the field $\phi(x,t)$
\begin{equation}\label{eq:eqmo}
\frac{\partial^2\phi}{\partial t^2} - \frac{\partial^2\phi}{\partial x^2} + \frac{dV}{d\phi} = 0,
\end{equation}
%
% For convenience, we are using natural coordinates and assume a previous scaling transformation $\phi \rightarrow \phi'$ which leaves the field adimensional.\\
where for the barrier case, the corresponding field equation is given by
\begin{equation}\label{Field Eq Barrier}
    \frac{\partial^2\phi}{\partial t^2}- \frac{\partial^2\phi}{\partial x^2}  + \phi - 4 \phi^3 \coth (2s) + 3 \phi^5 = 0,
\end{equation}
and for the well model we have
\begin{equation}\label{Field Eq Well}
    \begin{split}
         \frac{\partial^2\phi}{\partial t^2}- \frac{\partial^2\phi}{\partial x^2}  &+ \phi - \frac{3}{2} \phi^2 \left(2 + \cosh (2s) \right) + \phi^3 \left( \frac{7}{2} + 3 \cosh (2s) \right)\\
        &- \frac{15\phi^4}{4 \left( 1 - \tanh^2 s \right)}  + \frac{3\phi^5}{4 \left( 1 - \tanh^2 s \right)} = 0.
    \end{split}
\end{equation}
The static solutions that minimize the energy can be reduced to the first order ordinary differential equation
\begin{equation}\label{eq:eqmoStat}
    \partial_x \phi = \pm \sqrt{2 V(\phi,s)}.
\end{equation}
For topological or kink-like solutions, which are monotonic functions of $x$, the gradient does not change sign. This means that in Eq.~(\ref{eq:eqmoStat}), each sign corresponds to a distinct equation. For the positive sign, the solutions are known as kinks, and for the negative sign, they are called antikinks. However, non-topological or sphaleron-like solutions are not monotonic, and their gradient changes sign at the center of the solution, $x_0$. Although the potentials can take negative values, the sphaleron trajectory only probes the region
$V(\phi)\geq 0$. Starting from the false vacuum, for which \(V=0\), the field reaches the first zero of \(V\), where $\partial_x \phi = 0,$ and then returns to the same false vacuum. The sign of the square root therefore changes at the center $x=x_0$, as specified explicitly below. Therefore, for the above equation we have 
\begin{equation}\label{eq:SphaleronStat1}
\frac{d\phi}{d x}=\sqrt{2V(\phi)} \quad  x<x_0, \quad \frac{d\phi}{d x}=-\sqrt{2V(\phi)}  \quad  x>x_0.
\end{equation}
which corresponds to a bright sphaleron and for a dark sphaleron we have
\begin{equation}\label{eq:SphaleronStat2}
\frac{d\phi}{d x}=-\sqrt{2V(\phi)} \quad  x<x_0, \quad \frac{d\phi}{d x}=\sqrt{2V(\phi)}  \quad  x>x_0.
\end{equation}

Although the potentials in Eqs.~(\ref{eq:potentialB}) and (\ref{eq:potentialW}) do not possess
adjacent vacua with equal energy, the barrier model admits
a stable topological kink interpolating between its two
degenerate true vacua. The intermediate
false vacuum at $\phi=0$ does not prevent the existence of
such a solution, since its potential energy is higher than
that of the two true vacua. In contrast, the well model
does not admit an analogous static topological kink, since
its two degenerate false vacua at $\phi=0$ and $\phi=2$ are
separated by a true vacuum at $\phi=1$ with lower potential
energy. In the present work, however, we focus on
non-topological solitary-wave solutions, namely sphalerons,
which approach the same false vacuum at both spatial
infinities. Using the potentials in Eqs.~(\ref{eq:potentialB}) and (\ref{eq:potentialW}) and integrating equation (\ref{eq:eqmoStat}) with respect to $x$, one obtains

\begin{align}
    \phi_B(x,s) &= \pm \sqrt{\frac{\sinh(2s)}{\cosh(2s) + \cosh(2x)}}\label{barrier sphaleron},\\
    \phi_W(x,s) &= 1 \mp \frac{2 \sinh s \cosh\left( x / 2 \right)}{\sqrt{3 + \cosh(2s) + 2 \cosh x \sinh^2 s}}\label{well sphaleron}.
\end{align}
\begin{figure}
    \centering
    \includegraphics[width=\linewidth]{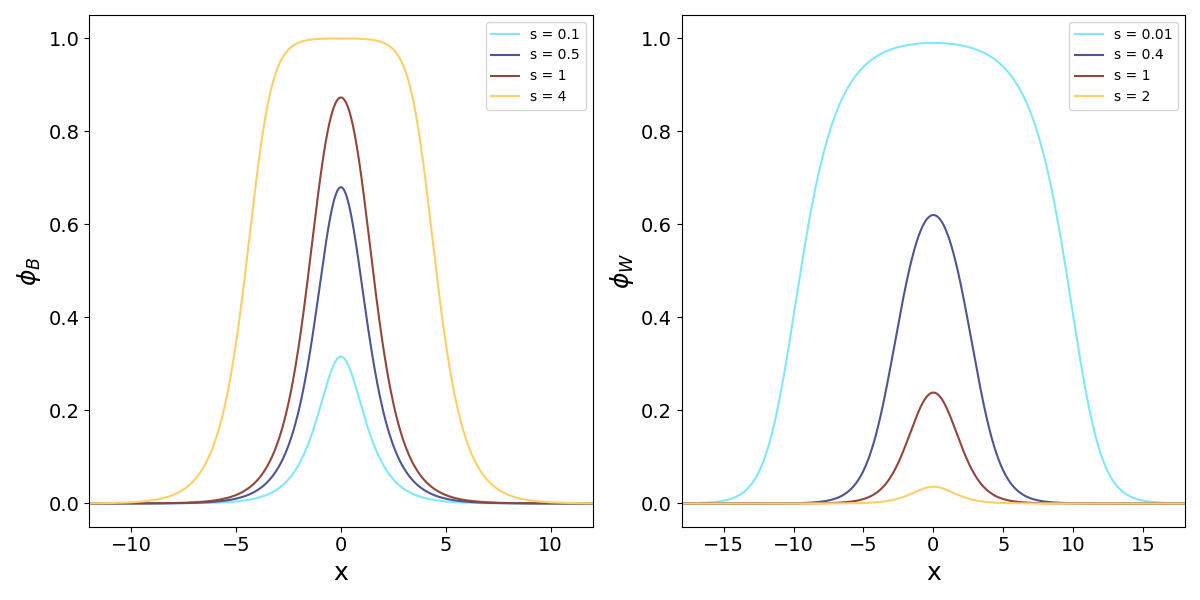}
    \caption{Bright sphaleron profiles in the barrier model (left) and the well model (right) for different values of $s$.}
    \label{fig:Sphaleron}
\end{figure}
In these expressions, the $+$ sign in Eq.~\eqref{barrier sphaleron} and the $-$ sign in Eq.~\eqref{well sphaleron} define the bright sphalerons, whereas the opposite signs define the dark sphalerons. The profiles of the sphalerons are shown in Fig.~\ref{fig:Sphaleron}, with the left (right) panel corresponding to the barrier (well) model. It can be observed that the sphaleron height increases and it becomes wider with increasing (decreasing) $s$ in the barrier (well) model, and it resembles a kink–-antikink pair. However, for decreasing (increasing) $s$ in the barrier (well) model, the profile takes the form of a localized lump.

\section{Linear Instability}\label{Sec:LinST}
The mode structure of sphaleron solutions is obtained by linearizing the field equations \eqref{Field Eq Barrier} and \eqref{Field Eq Well}. After inserting the perturbed field $\phi(x,t)=\phi(x)+\epsilon\eta(x)\cos(\omega t)$ into these expressions and retaining terms up to first order in $\epsilon$, the linearized field equations reduce to a Sturm-Liouville problem
\begin{equation}
    \left( -\partial_x^2 + U\left( \phi \right) \right) \eta(x) = \omega^2 \eta(x),
\end{equation}
where $U=U(x)$ represents the quantum mechanical potential
\begin{equation}\label{eq:QuantumMechanical}
U(x)=\frac{d^2V}{d\phi^2}\bigg|_{\phi(x)}.
\end{equation}

\begin{figure}
    \centering
    \includegraphics[width=\linewidth]{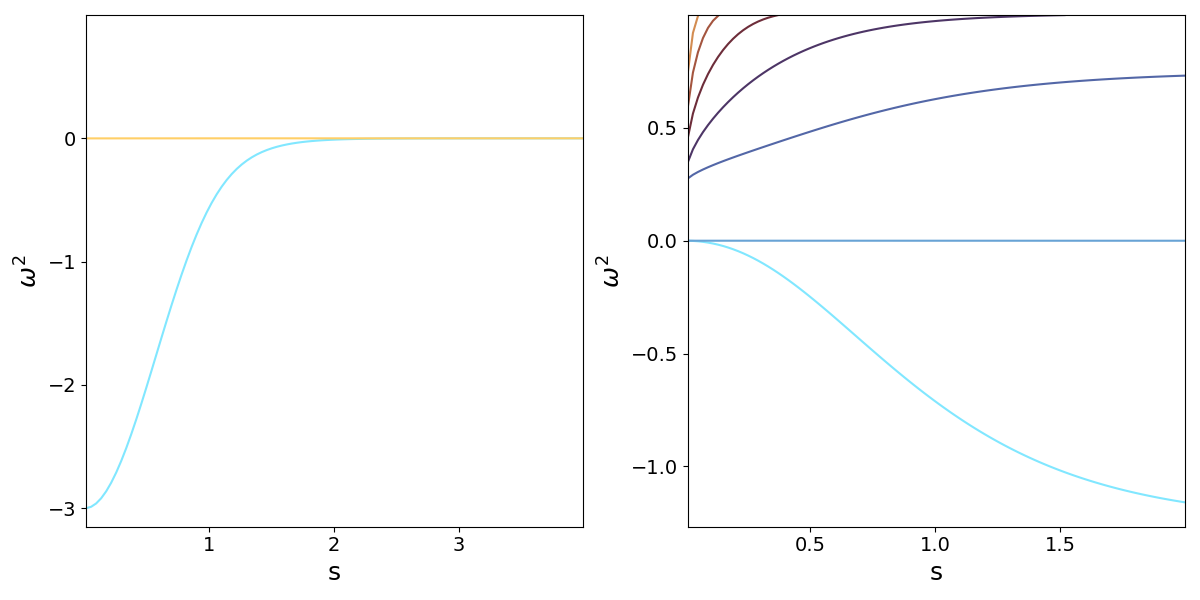}
    \caption{Eigenvalues for the bound, zero and unstable modes in the barrier (left) and well (right) models. In both cases the continuum threshold is at $\omega^2 = 1$.}
    \label{Eigenvalues}
\end{figure}

By studying the sphaleron excitation spectrum in the presence of the quantum mechanical potential $U(x)$, one can investigate how the potential energy  influences the dynamics and behavior of the sphaleron. The potentials $U(x)$ for sphalerons of the barrier and well models can be obtained by substituting Eqs.~(\ref{eq:potentialB}) and (\ref{eq:potentialW}) in Eq.~(\ref{eq:QuantumMechanical}) at $\phi=\phi_{B(W)}(x)$
\begin{align}
   U_{B} &= 1 - 6 \phi_B^2 (\tanh s + \coth s) + 15 \phi_B^4,\\
    U_{W} &= \frac{3}{1 - \tanh^2 s} \left( \frac{1 - \tanh^2 s}{3} - \left( 3 - \tanh^2 s \right)\phi_W + \left( 13 - \tanh^2 s \right) \phi_W^2 - 5 \phi_W^3 + \frac{5}{4} \phi_W^4 \right).
\end{align}
\begin{figure}
    \centering
    \includegraphics[width=\linewidth]{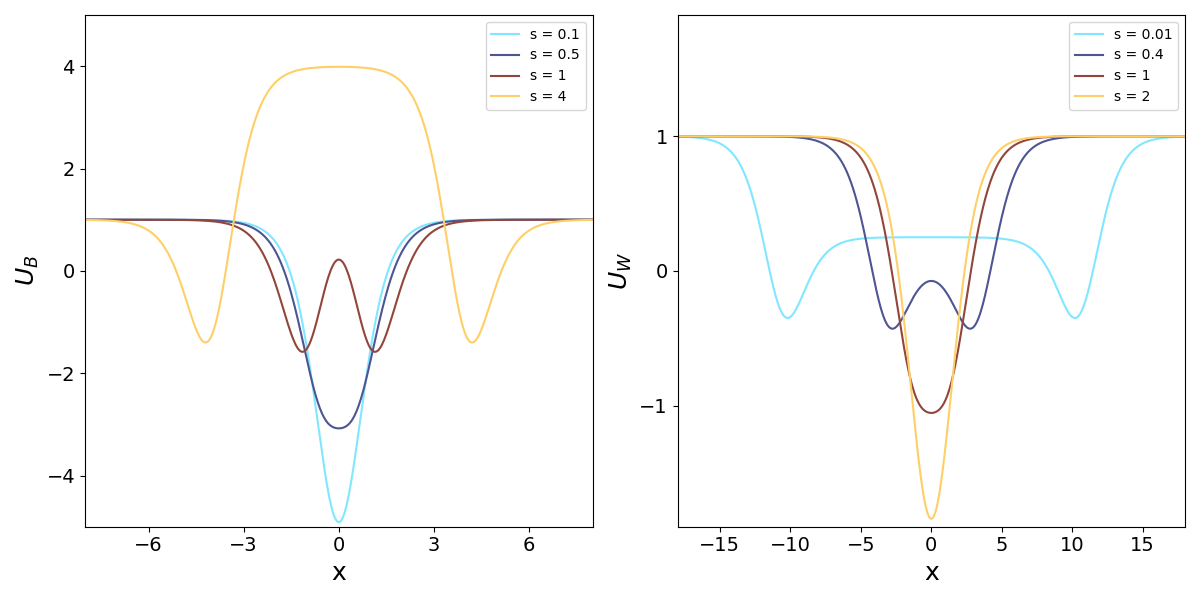}
    \caption{Quantum potentials for the sphalerons represented in Fig. \ref{fig:Sphaleron}}
    \label{fig:Quantum potentials}
\end{figure}
Fig.~\ref{fig:Quantum potentials} shows the quantum mechanical potentials for the aforementioned models. In the left panel, which corresponds to the barrier model, the effective potential develops a barrier at its center for large values of $s$. This can be explained by the fact that the sphaleron resembles a kink--antikink pair in the standard $\phi^6$ model at this limit. Such a pair is known not to possess a shape mode \cite{Dorey:2011yw}. The spectrum of the barrier model is shown in Fig.~\ref{Eigenvalues} as a function of the parameter $s$. It can be seen that the linear spectrum contains no bound states other than the zero mode and the unstable negative mode. The magnitude of the unstable mode increases with increasing $s$, indicating that the sphalerons become more stable for larger values of $s$. In contrast, for small values of $s$, they are highly unstable, and the height of the sphaleron approaches that of the true vacuum at early time.

However, the spectrum of the well model is more complicated, as it contains several bound states in addition to the zero mode and the unstable mode. This is clearly illustrated in the right panel of Fig.~\ref{Eigenvalues}, where the number of bound states increases as $s$ decreases. This behavior can be explained by the fact that the corresponding effective potential resembles that of an antikink--kink pair in the standard $\phi^6$ model. In that model, the number of shape modes increases as the separation between the antikink and kink becomes larger, as shown in the right panel of Fig.~\ref{fig:Quantum potentials}.

\begin{figure}
    \centering
    \includegraphics[width=\linewidth]{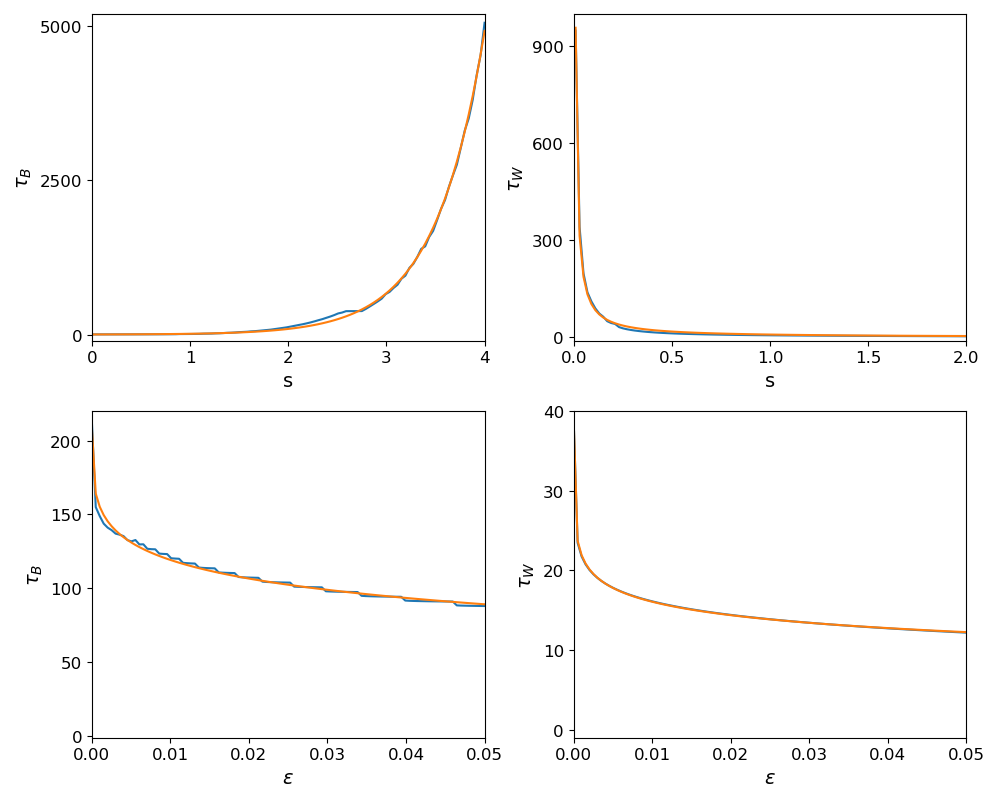}
    \caption{Lifetime of the sphalerons as a function of the deformation parameter $s$ (top panels) and the perturbation amplitude $\varepsilon$ (bottom panels). The left and right panels correspond to the barrier and well models, respectively. The blue curves represent the numerical results, while the orange curves show the corresponding fitted functions. The results demonstrate that the sphaleron lifetime increases with $s$ in the barrier model but decreases in the well model, whereas increasing the perturbation amplitude reduces the lifetime nonlinearly in both models.}
    \label{fig:lifetimes}
\end{figure}

To determine the lifetime of the sphalerons numerically, we perturb the static sphaleron configuration along its unstable eigenmode according to
\begin{equation}\label{eq:lifetime}
\phi(x;s) = \phi_{\mathrm{B(W)}}(x;s)+\varepsilon \eta_{-1}(x),
\end{equation}
where $\varepsilon$ is the perturbation amplitude, which is taken to be $\varepsilon=0.01$ throughout our simulations for the lifetime of the sphalerons as a function of the deformation parameter $s$, and $\eta_{-1}(x)$ is the normalized eigenfunction corresponding to the negative eigenvalue of the linear stability operator. It is clear that increasing the perturbation amplitude reduces the sphaleron lifetime by driving the system more rapidly away from the unstable equilibrium configuration. Since the sphaleron is unstable, even a small perturbation causes it to depart from the static configuration and eventually decay to kink--antikink for a positive perturbation parameter, allowing its lifetime to be determined from the subsequent time evolution. To determine the lifetime of the sphalerons, we measure the time required for the field at the center of the sphaleron to depart from its initial configuration and reach the value of the true vacuum. The resulting lifetimes for both the barrier and well models are presented in Fig.~\ref{fig:lifetimes}. In the top-left panel, we show the results for the barrier model, where the numerical data are fitted by an exponential function. We find that the sphaleron lifetime increases exponentially in the barrier model, approximately as $\tau_{B}\sim e^{2s}$. In contrast, for the well model (top-right panel), the lifetime follows an inverse power-law behavior, $\tau_{W}\sim s^{-1}$, indicating that the dependence of the sphaleron lifetime on the model parameter is qualitatively different in the two cases. Therefore, increasing $s$ enhances the stability of the sphaleron configurations in the barrier model, whereas it accelerates their decay in the well model. This contrasting behavior reflects the different effects of the deformation on the local structure of the potential around the sphaleron solution and its unstable mode.

To investigate the dependence of the sphaleron lifetime
on the initial perturbation amplitude, we have repeated
the numerical calculations for different values of
$\varepsilon$ in both the barrier and well models.
The corresponding results for two selected examples, $s = 2$ for the barrier model and $s = 0.4$ for the well model, are presented in the bottom
panels of Fig.~\ref{fig:lifetimes}. In both models, the lifetime decreases
nonlinearly with increasing perturbation amplitude.
Within the investigated range, the numerical results
can be described by the empirical fitting functions
$\tau_{B}\sim e^{-1.86\varepsilon^{0.23}}$ and $\tau_{w}\sim e^{-2.29\varepsilon^{0.11}}$ for barrier and well models, respectively. These fits are represented by the orange curves in the figure. These results indicate a qualitatively similar dependence
on the perturbation amplitude in the two models,
despite their different dependence on the deformation
parameter. We emphasize that these expressions are
empirical fits valid over the investigated range of
nonzero perturbations. In the limit of sufficiently small
perturbations, linear instability theory instead predicts
a logarithmic increase in the life time as
$\epsilon$ approaches zero. Therefore, the lifetime
should be interpreted as an operational quantity
depending on the initial perturbation rather than an
intrinsic property of the sphaleron.

It is important to emphasize that the lifetime analysis
presented in this section is performed independently
of the scattering simulations discussed below. Its
main purpose is to provide an approximate numerical
estimate of the time scale over which the sphalerons
retain their shape before the intrinsic instability
becomes significant. These estimates help to identify
the regions of parameter space in which sphalerons
can collide before undergoing spontaneous decay.
In particular, the scattering simulations presented
in Sec.~\ref{SphaleronScat} are performed without introducing any
explicit perturbation along the unstable eigenmode.
The perturbations considered in the present section
are therefore not part of the initial conditions
used for those scattering simulations. Explicit
perturbations are introduced separately in
Sec.~\ref{SphaleronScatPerturb} to investigate the subsequent interactions of bubble configurations produced
by sphaleron decay.

\section{Scattering of Unperturbed Sphalerons }\label{SphaleronScat}

In this section, we investigate the collision between two bright sphalerons in the barrier and well models given by Eqs.~(\ref{barrier sphaleron}) and (\ref{well sphaleron}). We set the initial conditions as follows: given a value of $s$, we place the first sphaleron $\phi_{B(W)}(x-x_1-v_1t)$ and another sphaleron $\phi_{B(W)}(x+x_1+v_1t)$ at positions $x_1 = 5\sigma_{B(W)}(s)$ and $x_2 = -x_1$, respectively. 
Here, $\sigma_{B(W)}(s)$ denotes the standard deviation of the sphaleron profile, treating $\phi(x)$ as a distribution over the spatial coordinate $x$, such that
\begin{equation}
    \sigma_\mathrm{B(W)}^2 = \int_{-\infty}^\infty x^2\hat{\phi}_\mathrm{B(W)}dx - \left(\int_{-\infty}^\infty x\hat{\phi}_\mathrm{B(W)}dx\right)^2,\quad\hat{\phi}_\mathrm{B(W)} = \frac{\phi_\mathrm{B(W)}}{\int_{-\infty}^\infty \phi_\mathrm{B(W)}dx}.
\end{equation}
This allows the initial separation between two sphalerons to be scaled according to their physical size, ensuring comparable initial conditions for all values of the deformation parameter $s$. The initial velocities are $v_1 = -v_0$ and $v_2 = v_0$. The model does not have an exact two-sphaleron solution; therefore, we utilize the following initial configuration
\begin{equation}\label{SphaleronInitial}
\phi_{\mathrm{init}}(x,t;s) = \phi_{\mathrm{B(W)}}(x-x_1-v_1t;s)+\phi_{\mathrm{B(W)}}(x+x_1+v_1t;s).
\end{equation}
This is an approximate solution for $t=0$ to some $t=t^{\star}$ just before the two profiles start to interact. Here $t^{\star}$ is the time of collision when the peaks would reach $x=0$. The initial data used in the numerical runs is
\begin{equation}\label{InitialData}
    \begin{split}
        \phi(x,0;s) &= \phi_{\mathrm{B(W)}}(x-x_1;s)+\phi_{\mathrm{B(W)}}(x+x_1;s),\\
        \phi_{\mathrm{t}}(x,0;s) &= -v_1\partial_x\phi_{\mathrm{B(W)}}(x-x_1;s)+v_1\partial_x\phi_{\mathrm{B(W)}}(x+x_1;s).
    \end{split}
\end{equation}

In most numerical simulations presented in this section, we use a spatial grid with $12001$ points, corresponding to an $x$ range of  $-200$ to $200$ with separation $h = 1/30$. Additionally, absorbing boundary conditions are applied to prevent reflections of small-amplitude radiation from the boundaries. We employ a 4th-order finite difference method to discretize the term $\phi_{xx}$ in Eq.(\ref{eq:eqmo}), minimizing the impact of discretization errors. For the time evolution, we employed the 4th-order St\o{}rmer-Verlet method with a time step of $\tau= 0.0025$.\\

In the following subsections, we present the results of the sphaleron scattering simulations for each model separately.

\subsection{Scattering in the barrier model}
\begin{figure}
    \centering
    \includegraphics[width=\linewidth]{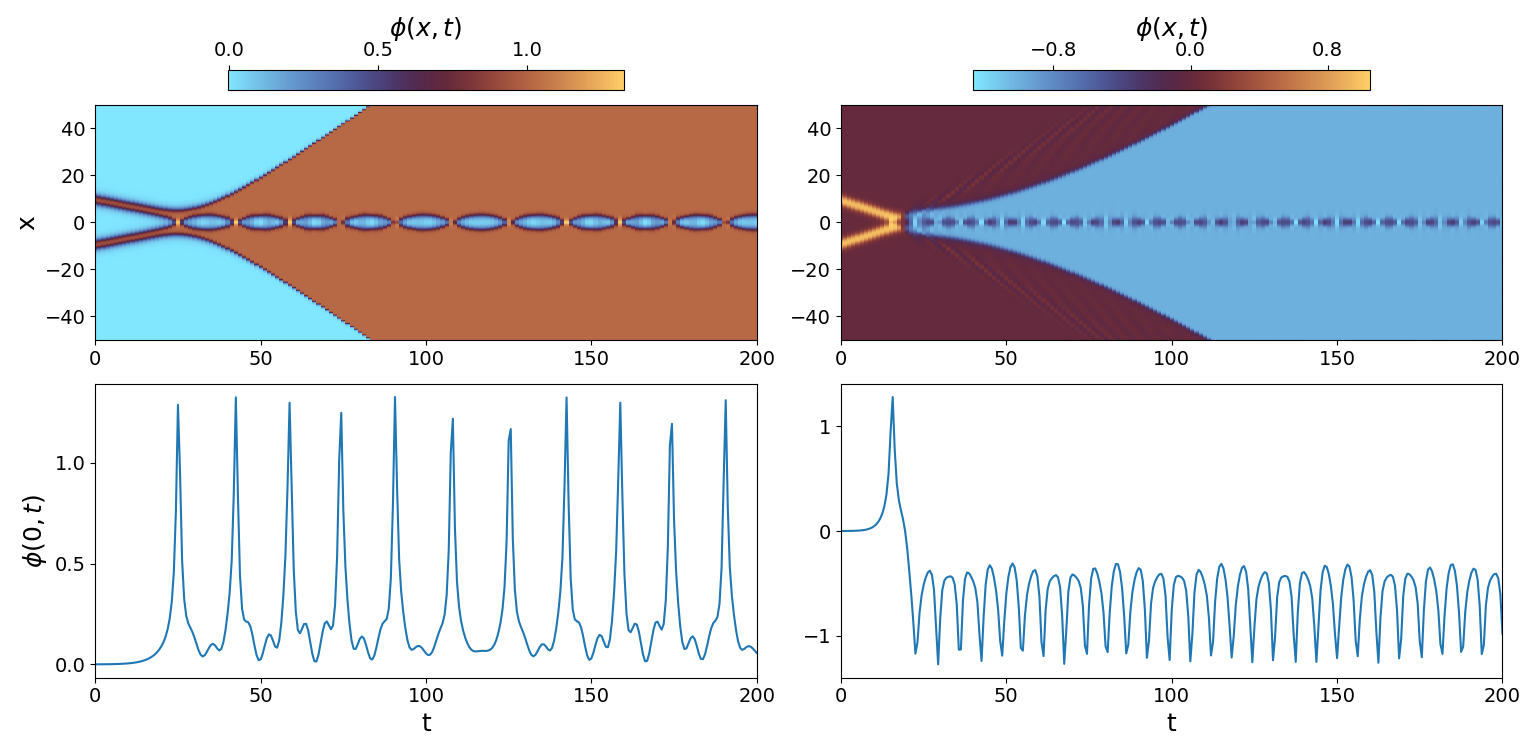}
    \caption{Top panels: Oscillons produced in the true vacua of the barrier model, for parameters $s = 1.2$, $v = 0.27$ (left) and $s = 1.6$, $v = 0.45$ (right). Bottom panels: Time evolution of the field at the collision point, $\phi(0,t)$, used to determine the oscillation period of the oscillon and the corresponding frequency.}
    \label{fig:expanding vacua barrier}
\end{figure}
\begin{figure*}[ht!] 
\begin{center}
  \centering
%  \subfigure[]
  {\includegraphics[width=\linewidth]{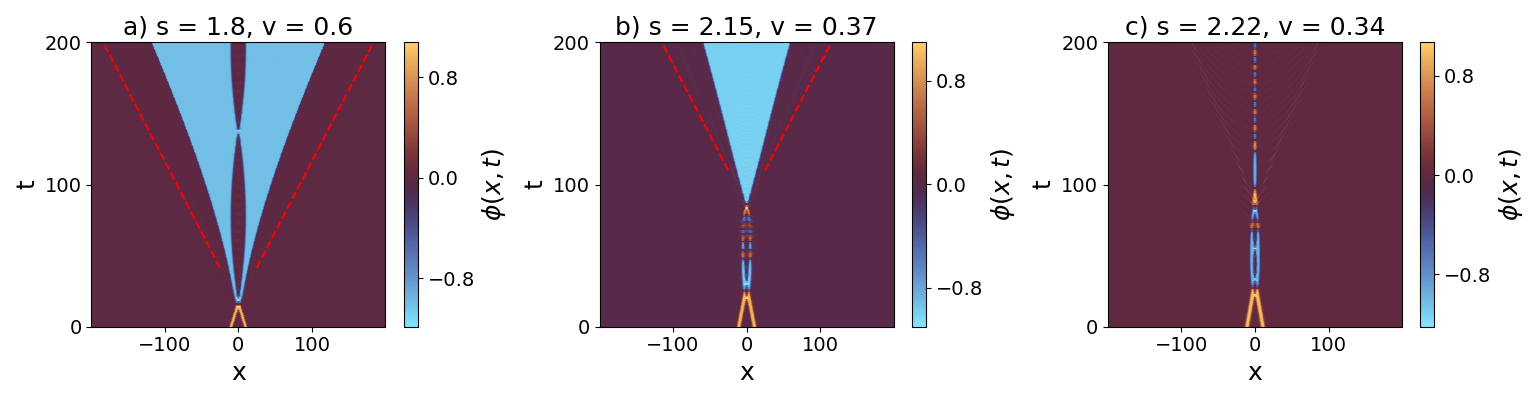}}
%    \subfigure[]
    {\includegraphics[width=\textwidth]{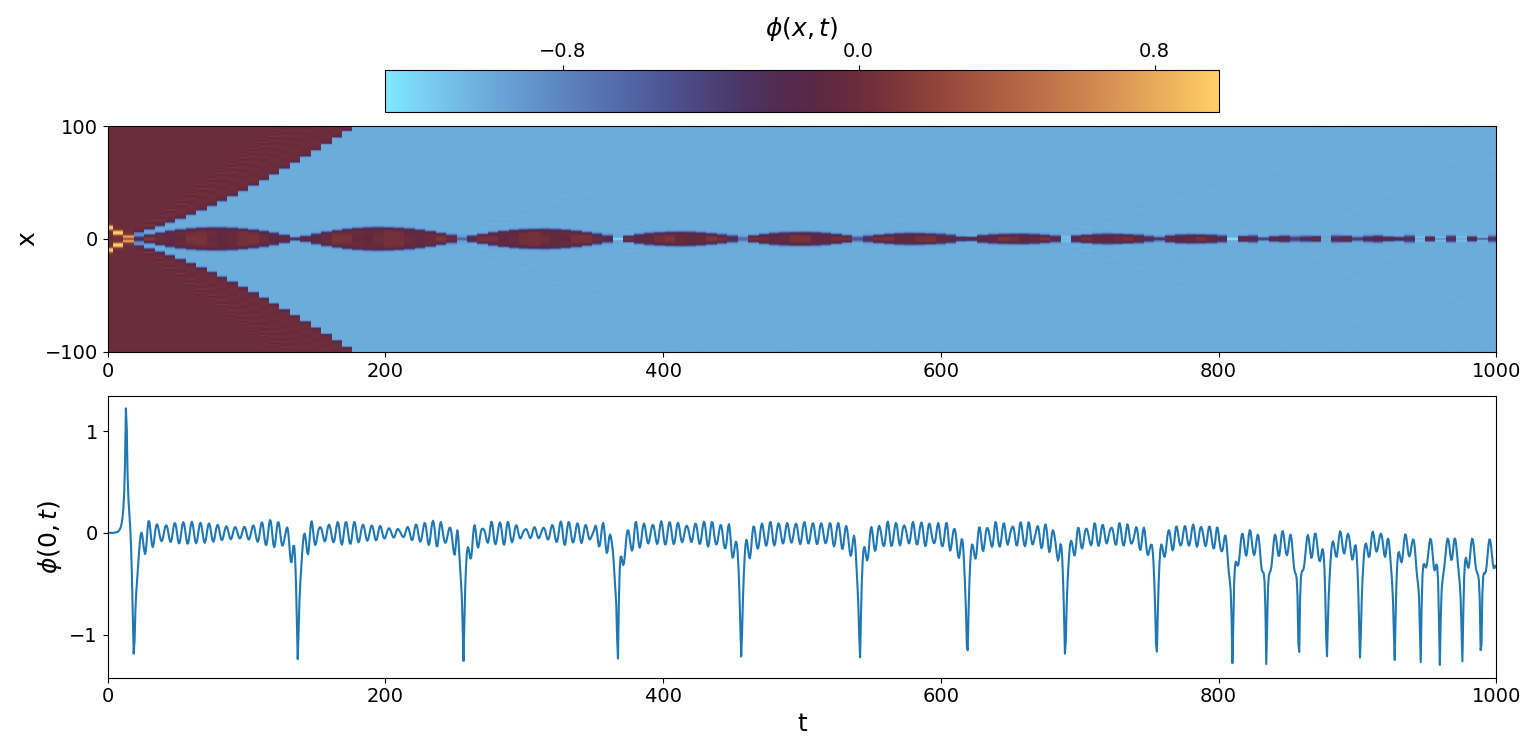}}
    \\
  \caption{Top panels: Bubble, kink--antikink pair and single oscillon in the final states of the sphaleron collision in a barrier model. The red dashed line indicates the light-cone boundary. Bottom panel: Large time evolution of the field at the collision point, $\phi(0,t)$, illustrating the formation and subsequent decay of the bubble.}
\label{fig:others barrier}
\end{center}
\end{figure*}

We begin by investigating the scattering of bright sphalerons in the barrier model with a false vacuum for different values of the parameter $s$. The sphalerons are initially placed far apart and boosted toward each other with various initial velocities, leading to a head-on collision at the origin. The resulting collision dynamics for several values of $s$ and initial velocities are illustrated in Figs.~\ref{fig:expanding vacua barrier}--\ref{fig:others barrier}. Three main scattering channels are observed in this process:
\begin{figure}
    \centering
    \includegraphics[width=\linewidth]{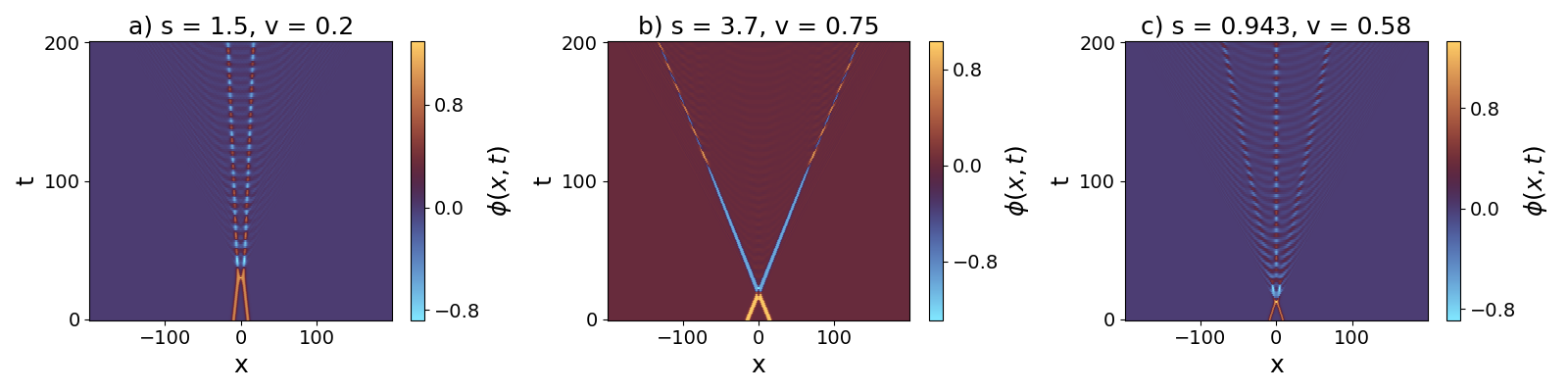}
    \caption{Oscillon pairs produced in the false vacuum of the barrier model. In the three-oscillon case, the central oscillon has an oscillation frequency of $\omega=0.86$, while the two outward-moving oscillons have $\omega=0.92$ after correcting for relativistic time dilation.}
    \label{fig:oscillons barrier}
\end{figure}

(i) The collision leads to the creation of kink--antikink pairs, which accelerate and approach the speed of light at large times, together with a long-lived oscillon localized at the collision center in the true vacuum region. To characterize the oscillon, we computed the time evolution of the field at the collision point, $\phi(0,t)$, and numerically determined its oscillation period by averaging over ten consecutive oscillations (see the bottom panels of Fig.~\ref{fig:expanding vacua barrier}). The corresponding oscillon frequency was then obtained from $\omega=2\pi/T$. We find $T\approx16.56$, yielding $\omega\approx0.379$ for the left panel, and $T\approx6.38$, yielding $\omega\approx0.986$ for the right panel. In both cases, the oscillation frequency lies below the mass threshold, $\omega=1$, which is a characteristic signature of long-lived oscillon configurations, as it suppresses efficient coupling to propagating linear radiation modes.

(ii) The collision results in the transformation of the sphalerons into one, two, or three oscillons in the false vacuum, which subsequently propagate away from the collision region, as shown in Fig.~\ref{fig:oscillons barrier} and Fig.~\ref{fig:others barrier}(c). A particularly interesting scenario is displayed in Fig.~\ref{fig:oscillons barrier}(b), where the bright sphalerons collide to produce a pair of dark sphalerons. These intermediate metastable configurations travel apart for some time before each decays into an oscillon, ultimately yielding two oscillons propagating in opposite directions. The dark sphalerons produced after the collision have a lifetime of approximately $t\approx 57$ (accounting for time dilation), which is considerably shorter than that of an isolated sphaleron perturbed along its unstable eigenmode according to Eq.~(\ref{eq:lifetime}). This difference indicates that the post-collision environment, including the presence of radiation and nonlinear interactions with the surrounding field, can significantly modify the decay rate of the metastable sphaleron.

(iii) The formation of a single kink--antikink pair accompanied by continuous radiation emission, as shown in Fig.~\ref{fig:others barrier}(b). Here, at the early stages of the collision, oscillons are produced as intermediate states. These oscillons can subsequently merge into a larger, long-lived oscillon configuration, which then acquires sufficient energy to overcome the sphaleron potential barrier and decay into the true vacuum through the formation of a kink--antikink pair. A similar oscillon-mediated transition into kink--antikink pair has recently been observed in related scalar field models \cite{campos2026collisiondynamicsfalsevacuumoscillons}.
\begin{figure}
    \centering
    \includegraphics[width=0.7\textwidth]{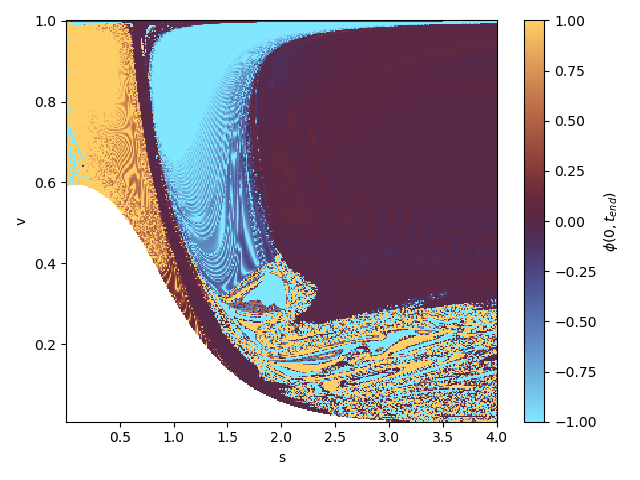}
    \caption{Phase diagram for the barrier model, showing the value of the field at the origin when $t = t_\mathrm{end} = \frac{x_0(s)}{v} + 169.78$. }
    \label{fig:Phase diagram barrier}
\end{figure}

In Fig.~\ref{fig:others barrier}(a), an interesting phenomenon is observed after the collision of the sphalerons: a long‑lived bubble of the false broken vacuum, bounded by a kink–antikink pair which accelerates and approaches the speed of light at large times. The red dashed line indicates the light-cone boundary. To confirm the nature of the intermediate state observed in Fig.~\ref{fig:others barrier}(a), we plot the space-time profile of the field  for a large time together with value of the filed at the origin, $\phi(x=0)$, as a function of time (see bottom panel). To account for the longer simulation time, an extended spatial grid from $x = -1200$ to $x = 1200$ was employed, keeping the separation of 1/30 for a total of 72001 points. The profiles clearly show the formation of a finite domain in which the field remains close to false  vacuum value, while the exterior approaches the other vacuum. These two regions are separated by a kink and an antikink, whose positions evolve in time as they move toward each other, collide, and subsequently re-emerge. This demonstrates that the intermediate configuration is indeed a vacuum bubble that later transforms into a localized oscillon. The repeated collapse and re-expansion of the bubble, accompanied by successive kink--antikink collisions, is characteristic of a bubble state driven by the energy difference between the two vacua (see our YouTube video: https://youtu.be/BhL4wvq1ELU).

The oscillating false-vacuum bubbles observed in our
simulations may be closely related to the large-amplitude
flat-top oscillons previously investigated in
Refs.~\cite{Amin:2010jq,Correa:2012gbo}.
These configurations share the presence of an extended,
nearly homogeneous spatial region. However, an important
feature of the configurations observed here is that this
region corresponds to a metastable vacuum bounded by
moving kink--antikink-like interfaces. Their evolution
is characterized by repeated contractions and expansions
of the vacuum domain, accompanied by successive
interactions between the interfaces. As energy is lost
through radiation, the bubble gradually becomes narrower
and eventually evolves into a localized oscillon.
This behavior suggests a possible connection between
oscillating vacuum bubbles and flat-top oscillons. Establishing whether these states belong to a continuously connected family would
require a more detailed investigation of their spatial
profiles, oscillation frequencies, and dependence on
the model parameters.

To provide a comprehensive picture of the scattering results, a phase diagram of the final states of the scattered sphaleron solutions for the barrier case with initial conditions (\ref{SphaleronInitial}) is presented in Fig~\ref{fig:Phase diagram barrier} at center of collision, i.e. $x=0$ at time $t=t_\mathrm{end}$, when the final objects have moved sufficiently far apart. Here, $t=t_{\mathrm{end}}$ denotes the time at which the field is evaluated at the collision point. This time is determined from the initial sphaleron position and its velocity, ensuring that no outgoing kink or radiative emission reaches the edges of the spatial grid. We have also excluded the parameter regions in which the sphalerons are highly unstable and decay before the interaction, these regions are shown as white areas in the phase diagrams. The left yellow region indicates the production of one pair of kink--antikink with a central oscillation, as shown in the example of Fig.\ref{fig:expanding vacua barrier} panel (a). Regions with pixels randomly distributed between dark blue and bright blue near the center of the image corresponds to the same phenomenon in the opposite true vacuum, shown in panels (b) of Fig. \ref{fig:expanding vacua barrier}. The bright blue regions indicate the creation of one pair of the kink--antikink along with some radiations, as depicted in Fig.~\ref{fig:others barrier} (b). The dark regions, which cover most of the phase diagram, indicate that the dominant outcome of the collision is the formation of a pair of oscillons in the final state. This mainly occurs for large values of $s$, where the energies of the true and false vacua become nearly equal, as shown in Fig.~\ref{fig:Potentials}(a). The low velocity region of the diagram is covered in a complex pattern of final states, as longer interaction times between the initial sphalerons allow for a greater variety of outcomes.

%In Fig.~\ref{fig:others barrier} (a), the initial colliders have an initial velocity of $v=0.95$, and an interesting phenomenon is observed after the collision of the sphalerons at the origin: One pair of kink-antikink is created and moves toward infinity, while a second pair of kink-antikink becomes trapped between them. The trapped pair moves toward each other, collides, moves apart, and then re-collides multiple times. This phenomenon was observed within a specific range of initial velocities, $0... \lesssim v \lesssim ...$, for the sphalerons. {\color{blue} We have to understand if this phenomeonon is bubble or something else happen here!!!}

\subsection{Scattering in the well model}
\begin{figure}
    \centering
    \includegraphics[width=\linewidth]{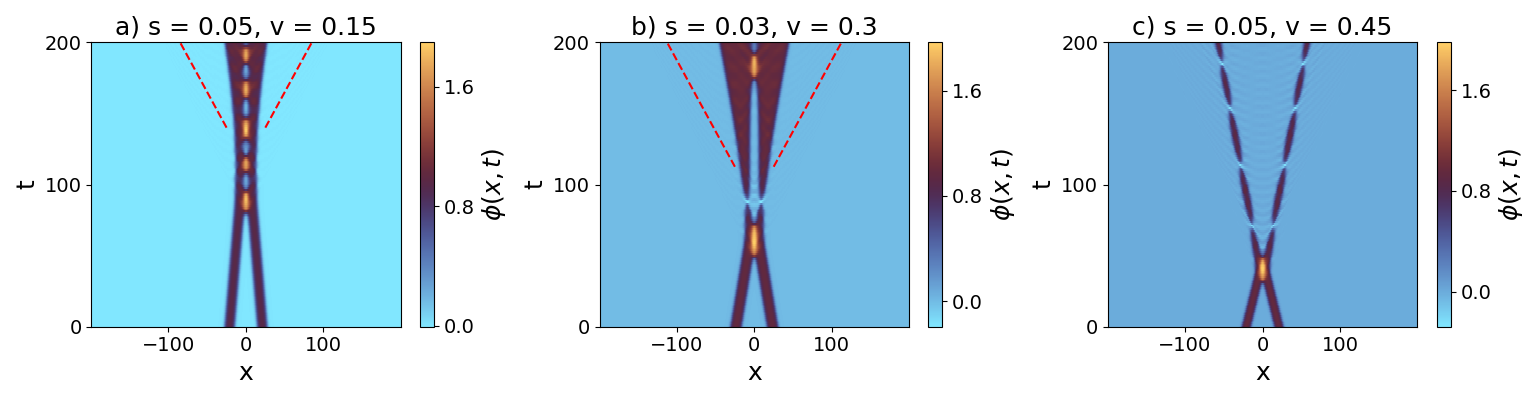}
    \caption{Panels (a) and (b): Formation of a kink--antikink pair together with an oscillon following the collision of two sphalerons. The red dashed line indicates the light-cone boundary. Panel (c): Formation of a pair of oscillons in the false vacuum of the well model.}
    \label{fig:low s well}
\end{figure}
\begin{figure*}[ht!] 
\begin{center}
  \centering
%  \subfigure[]
  {\includegraphics[width=\linewidth]{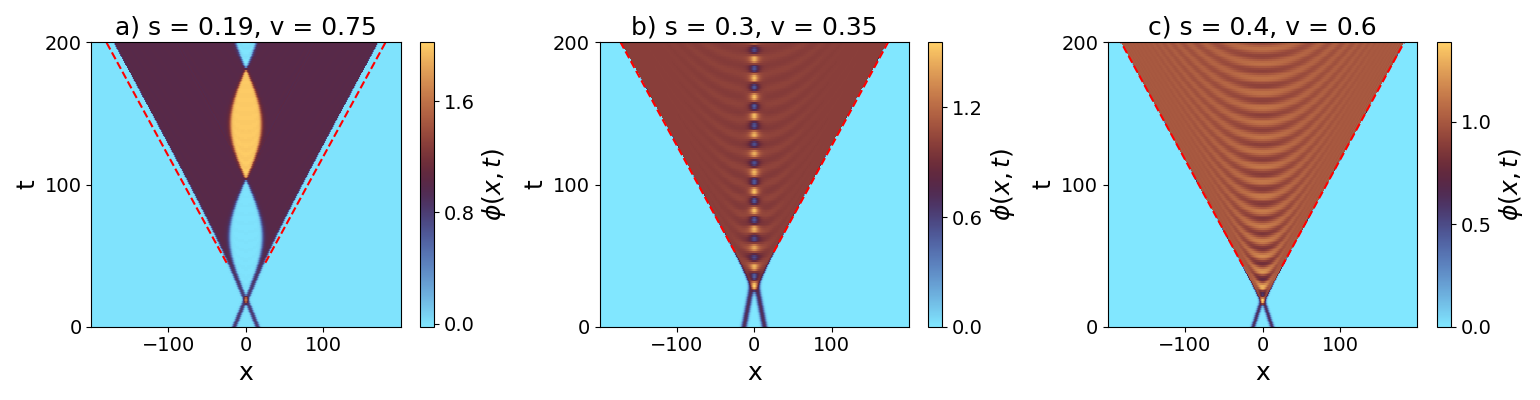}\label{fig:group_vel_Kinks_vel}}
%    \subfigure[]
    {\includegraphics[width=\textwidth]{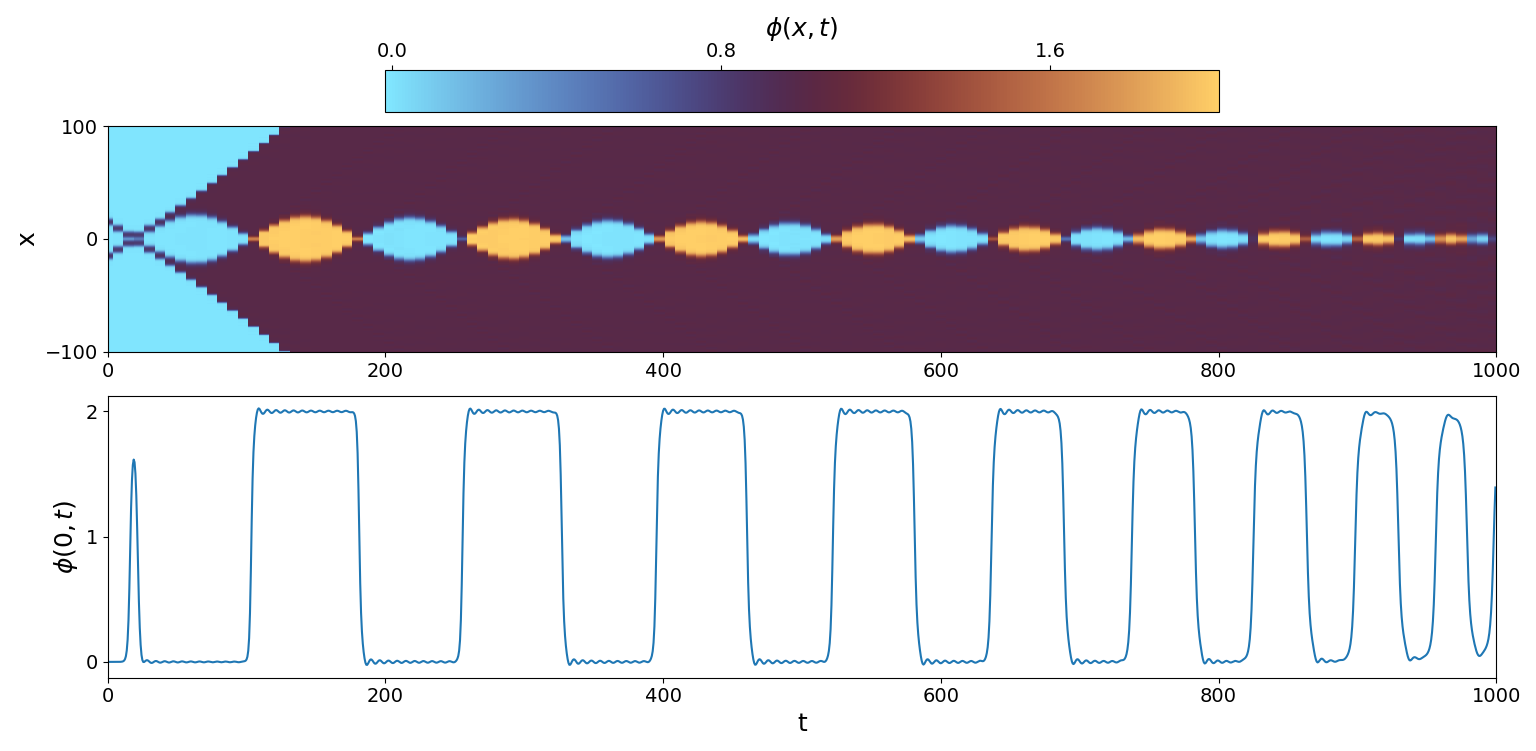}\label{fig:group_vel_shifted_vel}}
    \\
  \caption{Top panels: Final states of sphaleron collisions in the well model, showing the formation of bubbles and kink--antikink pairs accompanied by oscillons and radiation. The red dashed line indicates the light-cone boundary. Bottom panels: Long-time evolution of the field at the collision point, $\phi(0,t)$, illustrating the formation and subsequent decay of the transient bubble.}
\label{fig:others well}
\end{center}
\end{figure*}
Next, our focus is on studying the collision of two sphalerons in the well model with two false vacua. The outcomes are shown in Figs.~\ref{fig:low s well} and \ref{fig:others well} for various initial velocities and parameters of of the model. In Fig.~\ref{fig:low s well}(a) and (b), the two sphalerons move toward each other, collide at the origin, and transform into a pair of kink--antikink that accelerate to infinity with a localized oscillating mode that oscillates at the center. In (c), when the sphalerons collide, they transform into two oscillons in the false vacuum that move away to infinity. In Fig.~\ref{fig:others well}(a), we observe the same phenomenon as in Fig.~\ref{fig:others barrier}(a). In panel (b), the sphalerons decay into a kink--antikink pair that departs at high speed immediately after the collision, together with an oscillon in the false vacuum, while in (c), the final state consists of a kink--antikink pair accompanied by radiation.

In the bottom panel of Fig.~\ref{fig:others well}, we plot the time evaluation of the field at origin $\phi(0,t)$ for a longer time for the phenomenon that we observed in Fig.~\ref{fig:others well}(a), in a similar manner to the bottom panel of Fig.~\ref{fig:others barrier}. The snapshots reveal a region in which the field is frequently changing between both false vacua values of the well model, which is bounded by a kink and an antikink that repeatedly approach, collide, and separate. This confirms that the intermediate state is a vacuum bubble. The subsequent evolution consists of repeated collapses and re-expansions of the bubble, with the kink and antikink being recreated after each collision, giving rise to a long-lived bubble before its eventual decay into an oscillon (see our YouTube video: https://youtu.be/aEpV9lX85wA).

\begin{figure}
    \centering
    \includegraphics[width=0.7\textwidth]{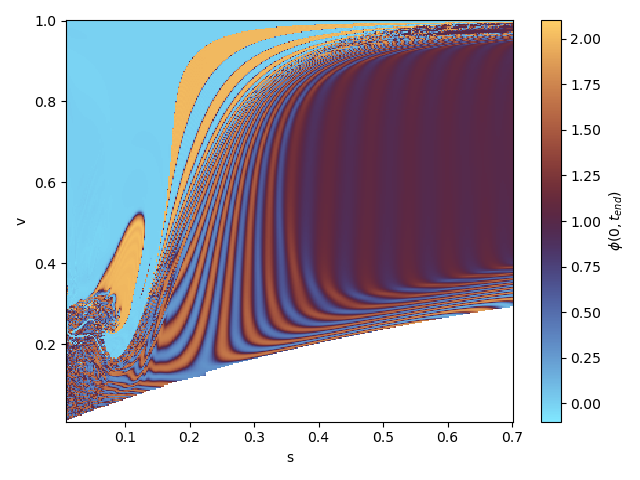}
    \caption{Phase diagram for the well model, showing the value of the field at the origin when $t = t_\mathrm{end} = \frac{x_0(s)}{v} + 170.70$.}
    \label{fig:Phase diagram well}
\end{figure}

A contour plot of the final states of the scattered sphalerons in the well model as a function of $v$ and $s$ is shown in the panel of Fig.\ref{fig:Phase diagram well}. In the dark region, only one pair of kink--antikink is found, accompanied by the emission of radiation in the final states. In the yellow region, the collision produces a false-vacuum bubble that undergoes a sequence of bounces, with kink--antikink pairs being re-created after each collision, as illustrated in Fig.~\ref{fig:others well}(a). The blue region corresponds either to the formation of two long-lived oscillons, as shown in Fig.~\ref{fig:low s well}(c), or to the temporary restoration of the vacuum phase during the bouncing dynamics described above. These outcomes reflect the rich nonlinear evolution of the scalar field following the collision and the competition between localized oscillatory states and repeated vacuum-domain formation. Regions with pixels randomly scattered between blue and yellow indicate an oscillatory pattern at $x=0$ accompanied by a pair of kink-antikink, as shown in Fig.~\ref{fig:low s well} (a,b). The graph shows that for a small $s$, the likelihood of producing two pairs of oscillons (blue region) increases. This once again corresponds to the scenario in which the energy levels of the true and false vacua are nearly degenerate, as illustrated in the right panel of Fig.~\ref{fig:Potentials}. The darker region covering the right half of the diagram corresponds to the formation of a kink--antikink pair with radiation, depicted in Fig.~\ref{fig:others well}.

\section{Scattering of Initially Static Perturbed Sphalerons}\label{SphaleronScatPerturb}

\begin{figure}
    \centering
    \includegraphics[width=\textwidth]{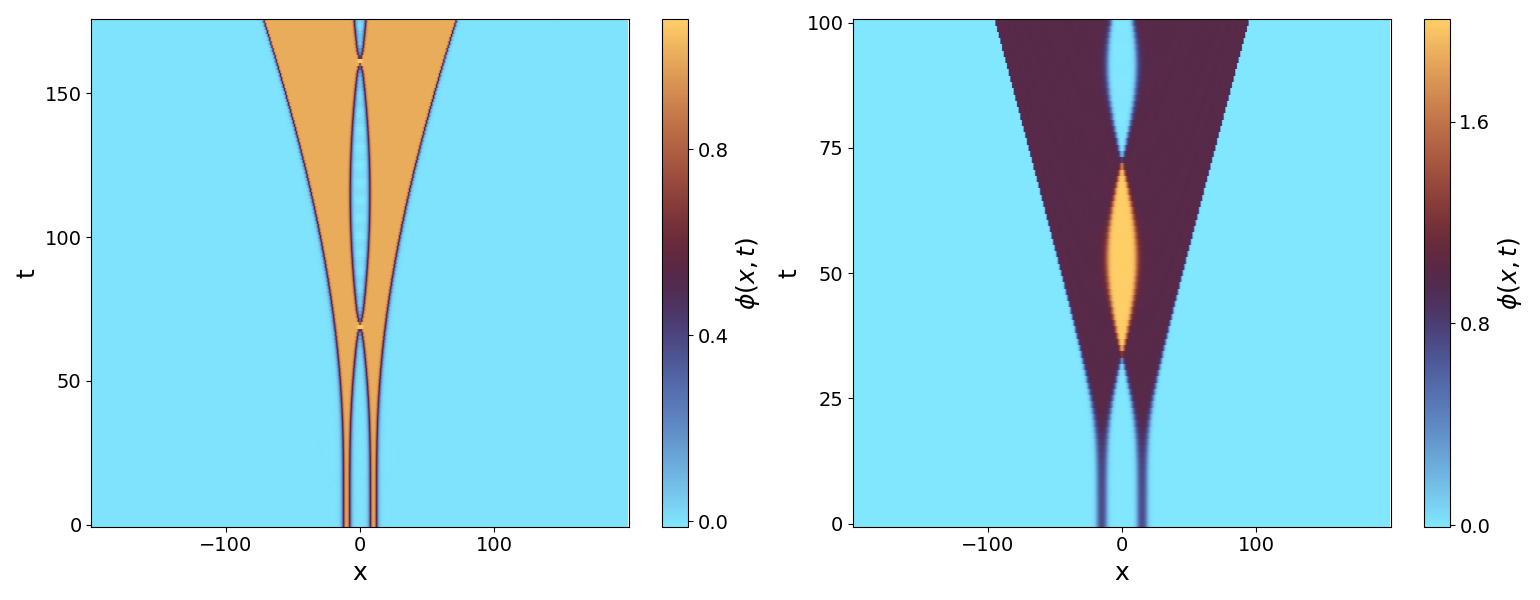}
    \caption{Scattering of initially static, perturbed sphalerons in the (a) barrier and (b) well models. The perturbation causes each sphaleron to decay into a kink--antikink pair before the collision.}
    \label{fig:kink scattering}
\end{figure}

We first investigate the scattering of two static sphalerons that are initially perturbed with the unstable eigenmode with a positive amplitude. As a result, each sphaleron decays into a kink--antikink pair before the two sphalerons collide. For the perturbation, we used a normalized gaussian function with the perturbation amplitude $\varepsilon=0.01$. The corresponding results for the barrier and well models are presented in Fig.~\ref{fig:kink scattering}, where the deformation parameters are chosen as $s=1.8$ and $s=0.3$, respectively.
It can be seen that each sphaleron splits into a kink--antikink pair. The two outer kinks accelerate away from the collision region, asymptotically approaching the speed of light, while the two inner kinks form a false-vacuum bubble. This bubble undergoes several oscillations (or bounces) before eventually collapsing into an oscillon. We have verified the robustness of this behavior by performing simulations for a wide range of initial separations and perturbation amplitudes. In all cases considered, a transient bubble is formed, indicating that its formation is a robust feature of the collision dynamics rather than a consequence of a particular choice of initial conditions. These results demonstrate that the decay products of unstable sphalerons undergo rich nonlinear interactions before eventually relaxing into long-lived localized configurations, such as oscillons.

\begin{figure}
    \centering
    \includegraphics[width=\textwidth]{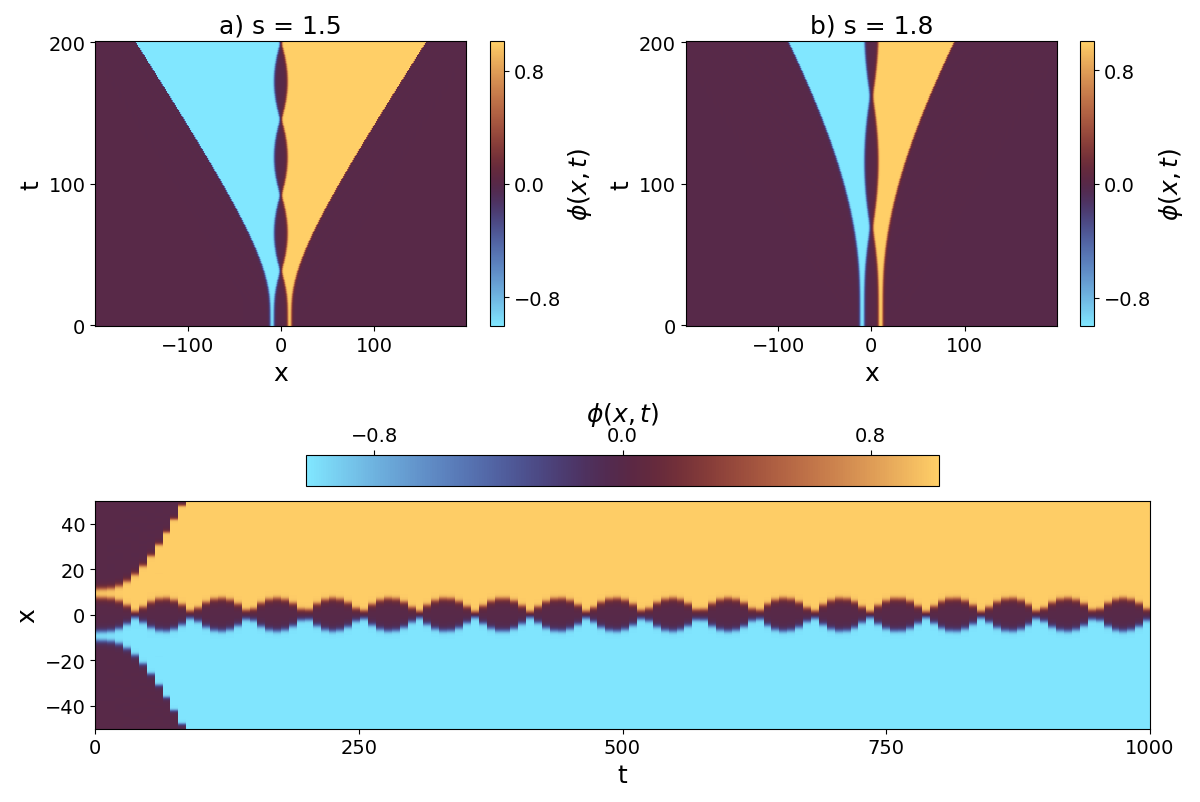}
    \caption{Top panels: Scattering of an initially static bright sphaleron and an initially static dark sphaleron in the barrier model after perturbation along their unstable eigenmodes. Bottom panels: Long-time evolution of the field, illustrating a long-lived false-vacuum bubble that persists throughout the simulation due to their mutual repulsive interaction.}
    \label{fig:bright dark}
\end{figure}

Figure~\ref{fig:bright dark} shows the scattering of an initially static bright sphaleron and an initially static dark sphaleron in the barrier model for different values of the deformation parameter $s$. Before the collision, both sphalerons are perturbed along their unstable eigenmodes, causing them to decay into kink--antikink pairs. The two outer antikinks accelerate away from the collision region and propagate toward spatial infinity, while the two inner kinks remain trapped and form a long-lived false-vacuum bubble. Unlike the bubble produced in the bright--bright sphaleron collisions, this bubble does not collapse into an oscillon even at late times. Instead, it remains stable over the duration of the simulation because the repulsive interaction between the two kinks prevents them from approaching each other and annihilating. This behavior is illustrated in the bottom panels of Fig.~\ref{fig:bright dark} (see our YouTube video: https://youtu.be/v4g9X5Shrys).

\begin{figure}
    \centering
    \includegraphics[width=\textwidth]{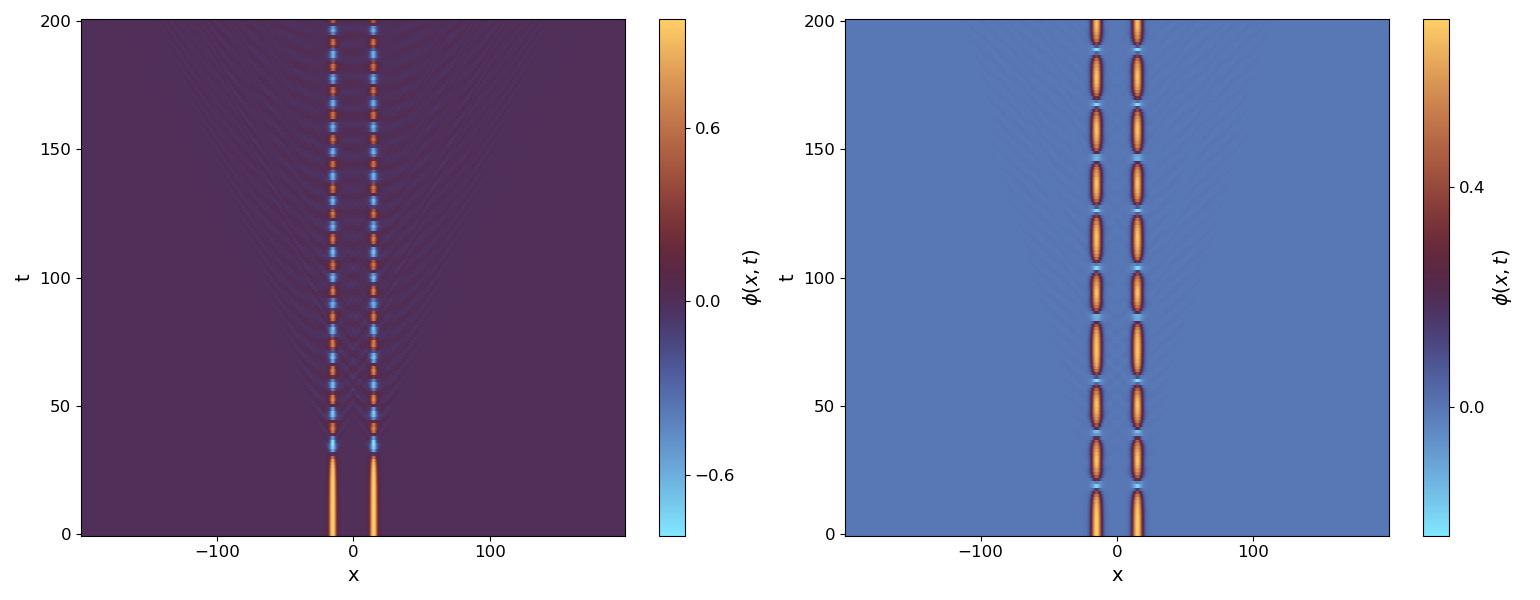}
    \caption{Evolution of two initially static sphalerons
under negative perturbations along their unstable
eigenmodes. The left and right panels correspond to
the barrier model with $s=1.8$ and the well model
with $s=0.3$, respectively. The perturbation amplitude
is $\epsilon=-0.01$ in both cases. The sphalerons
evolve into long-lived localized oscillatory
configurations that remain near their initial positions,
without producing expanding kink--antikink pairs
or false-vacuum bubbles during the simulated time
interval.}
    \label{fig:parallel Oscillon}
\end{figure}

To investigate the effect of the opposite sign of the
perturbation, we have also considered two initially
static sphalerons perturbed along the negative direction
of their unstable eigenmodes. The corresponding results
for the barrier and well models are presented in
Fig.~\ref{fig:parallel Oscillon}, using the same deformation parameters and
perturbation magnitude as in the positive-perturbation
simulations. In both models, negative perturbations
cause the sphalerons to decay into long-lived localized
oscillatory configurations, consistent with oscillon
formation. Unlike the positive-perturbation case,
where the sphalerons decay into false-vacuum bubble configurations, the resulting
oscillatory structures remain localized near their
initial positions.

Furthermore, we have repeated these simulations for
different values of the deformation parameter $s$
and perturbation amplitude $\epsilon$ in both models.
For all the negative-perturbation cases investigated,
we observed localized oscillatory configurations,
without the formation of false-vacuum bubbles.
These results demonstrate that, although the
sphalerons are unstable under both signs of the
perturbation, their nonlinear decay channels
depend strongly on the perturbation sign.
It should be emphasized that the present analysis
is restricted to initially static sphalerons.
The scattering dynamics of negatively perturbed
sphalerons with nonzero initial velocities may
exhibit different outcomes and constitute an
interesting direction for future investigation.

\section{Conclusions}\label{sec:conclusions}\label{Sec:Con}
In this work, we have presented an investigation of the collision dynamics of bright sphalerons in deformed $\phi^6$ models with symmetric potentials containing false vacua. We considered two complementary realizations of the model, namely the barrier model with one false vacuum and the well model with two false vacua, and examined how the interplay between the vacuum structure, the intrinsic instability of the sphalerons, and the linear excitation spectrum determines the nonlinear evolution following head-on collisions.

We first reviewed the linear stability properties of the sphalerons, emphasizing that the two models exhibit qualitatively different excitation spectra. The barrier model possesses only the translational mode together with the unstable mode, whereas the well model supports several additional localized bound states whose number increases as the deformation parameter is varied. However, the present analysis does not attempt to establish a quantitative relation between excitation of these bound modes and individual scattering channels; such an investigation would require a dedicated mode-projection analysis.

Extensive numerical simulations reveal a remarkably rich variety of final states. Depending on the deformation parameter and the initial collision velocity, the colliding sphalerons can decay into kink--antikink pairs, generate one or several long-lived oscillons or undergo combinations of these processes. Phase diagrams constructed over the parameter space provide a global classification of these regimes and demonstrate that small changes in the model parameters can lead to qualitatively different nonlinear evolution. In particular, oscillon production becomes increasingly prominent in regions where the energies of the true and false vacua are nearly degenerate, illustrating the important role played by the underlying vacuum landscape in determining the final state.

One of the most interesting results of this work is the observation of long-lived false-vacuum bubbles generated dynamically during sphaleron collisions. In these events, the collision produces a finite vacuum domain bounded by a kink--antikink pair that repeatedly collapses and re-expands before eventually decaying.  Unlike localized oscillons, these intermediate configurations represent oscillating vacuum domains whose time evolution cannot be governed by the local dynamics inside the false vacuum bubble. Instead, it is determined by the dynamics of the bubble walls, i.e., by the repeated kink--antikink interactions. To the best of our knowledge, such bubble formation has not previously been reported in real scalar field theories. Their appearance in both the barrier and well models suggests that transient vacuum-domain formation may be a robust feature of unstable localized defects evolving in theories with metastable vacua.

Several directions for future research naturally emerge from the present work. In particular, it would be of considerable interest to investigate sphaleron scattering in a $\phi^6$ model with a non-symmetric potential, where the three vacuum states are non-degenerate. Such a model is expected to exhibit a significantly richer landscape of collision dynamics, owing to the inequivalent vacua and the possibility of asymmetric vacuum transitions during the scattering process. Another promising direction is the investigation of oscillon--oscillon collisions in the barrier and well models considered in the present work. Given the diverse range of oscillon-mediated processes observed here, these collisions may reveal additional nonlinear phenomena, including resonance structures, crossing, and the possible formation of topological defects, analogous to the rich dynamics reported in \cite{campos2026collisiondynamicsfalsevacuumoscillons}.

\section*{Acknowledgments}

M.\@ A.\@ M.\@ S.\@ and D.\@ S.\@, are supported by the Higher Education and Science Committee of Armenia (HESCS), grant No. 25IRF/2-1C008.
C.\@ A. acknowledges financial support from the Spanish Research State Agency under project PID2023-152762NB-I00, the Xunta de Galicia under the project ED431F 2023/10 and the CIGUS Network of Research Centres, the María de Maeztu grant CEX2023-001318-M funded by MICIU/AEI /10.13039/501100011033, and the European Union ERDF. 

\bibliography{bibliography}

@article{Navarro-Obregon:2024ieb,
    author = "Navarro-Obreg{\'o}n, S. and Queiruga, J.",
    title = "{Impact of the internal modes on the sphaleron decay}",
    eprint = "2406.05706",
    archivePrefix = "arXiv",
    primaryClass = "hep-th",
    doi = "10.1140/epjc/s10052-024-13175-w",
    journal = "Eur. Phys. J. C",
    volume = "84",
    number = "8",
    pages = "821",
    year = "2024"
}

@article{Manton:1988az,
    author = "Manton, N. S. and Samols, T. M.",
    title = "{Sphaleron on a circle}",
    reportNumber = "DAMTP/88-5",
    doi = "10.1016/0370-2693(88)91412-8",
    journal = "Phys. Lett. B",
    volume = "207",
    pages = "179--184",
    year = "1988"
}

@article{MateosGuilarte:1992kil,
    author = "Mateos Guilarte, Juan",
    title = "{Sphalerons and instantons in two-dimensional field theory}",
    reportNumber = "PRINT-92-0065 (SALAMANCA)",
    doi = "10.1016/0003-4916(52)90044-4",
    journal = "Annals Phys.",
    volume = "216",
    pages = "122--151",
    year = "1992"
}

@article{Shnir:2009ct,
    author = "Shnir, Ya. and Tchrakian, D. H.",
    title = "{Skyrmion-Anti-Skyrmion Chains}",
    eprint = "0906.5583",
    archivePrefix = "arXiv",
    primaryClass = "hep-th",
    doi = "10.1088/1751-8113/43/2/025401",
    journal = "J. Phys. A",
    volume = "43",
    pages = "025401",
    year = "2010"
}

@article{Manton:2019qka,
    author = "Manton, N. S.",
    editor = "Dainton, John",
    title = "{The Inevitability of Sphalerons in Field Theory}",
    eprint = "1903.11573",
    archivePrefix = "arXiv",
    primaryClass = "hep-th",
    reportNumber = "DAMTP-2019-10",
    doi = "10.1098/rsta.2018.0327",
    journal = "Phil. Trans. Roy. Soc. Lond. A",
    volume = "377",
    number = "2161",
    pages = "20180327",
    year = "2019"
}

@article{Manton:2023mdr,
    author = "Manton, N. S. and Roma{\'n}czukiewicz, T.",
    title = "{Simplest oscillon and its sphaleron}",
    eprint = "2301.09660",
    archivePrefix = "arXiv",
    primaryClass = "hep-th",
    doi = "10.1103/PhysRevD.107.085012",
    journal = "Phys. Rev. D",
    volume = "107",
    number = "8",
    pages = "085012",
    year = "2023"
}

@article{Klinkhamer:1990ik,
    author = "Klinkhamer, Frans R.",
    title = "{A New Sphaleron in the {Weinberg-Salam} Theory}",
    reportNumber = "NIKHEF-H/90-9",
    doi = "10.1016/0370-2693(90)91319-7",
    journal = "Phys. Lett. B",
    volume = "246",
    pages = "131--134",
    year = "1990"
}

@article{Kunz:1988sx,
    author = "Kunz, J. and Brihaye, Y.",
    title = "{New Sphalerons in the {Weinberg-Salam} Theory}",
    doi = "10.1016/0370-2693(89)91130-1",
    journal = "Phys. Lett. B",
    volume = "216",
    pages = "353--359",
    year = "1989"
}

@article{Taubes:1982ie,
    author = "Taubes, Clifford Henry",
    title = "{The Existence of a Nonminimal Solution to the SU(2) {Yang-Mills} Higgs Equations on R**3}. Part I",
    reportNumber = "HUTMP-82-B118",
    doi = "10.1007/BF01206014",
    journal = "Commun. Math. Phys.",
    volume = "86",
    pages = "257",
    year = "1982"
}

@article{Taubes:1982ieII,
    author = "Taubes, Clifford Henry",
    title = "{The Existence of a Nonminimal Solution to the SU(2) {Yang-Mills} Higgs Equations on R**3}. Part II",
    doi = "10.1007/BF01212170",
    journal = "Commun. Math. Phys.",
    volume = "86",
    pages = "299-320",
    year = "1982"
}

@article{Saurabh:2019rrp,
    author = "Saurabh, Ayush and Vachaspati, Tanmay",
    editor = "Dainton, John",
    title = "{Monopole{\textendash}antimonopole: interaction, scattering and creation}",
    eprint = "1904.02257",
    archivePrefix = "arXiv",
    primaryClass = "hep-th",
    doi = "10.1098/rsta.2019.0143",
    journal = "Phil. Trans. Roy. Soc. Lond. A",
    volume = "377",
    number = "2161",
    pages = "20190143",
    year = "2019"
}

@article{Krusch:2004uf,
    author = "Krusch, Steffen and Sutcliffe, Paul",
    title = "{Sphalerons in the Skyrme model}",
    eprint = "hep-th/0407002",
    archivePrefix = "arXiv",
    doi = "10.1088/0305-4470/37/38/008",
    journal = "J. Phys. A",
    volume = "37",
    pages = "9037",
    year = "2004"
}

@article{Isler:1989tt,
    author = "Isler, K. and Letourneux, J. and Paranjape, Manu B.",
    title = "{Sphaleron in the skyrme model}",
    doi = "10.1103/PhysRevD.40.2490",
    journal = "Phys. Rev. D",
    volume = "40",
    pages = "2490--2492",
    year = "1989"
}

@article{Shnir:2011zza,
    author = "Shnir, Ya and Tchrakian, D. H.",
    editor = "Angelova, Maia and Zakrzewski, Wojciech and Hussin, Veronique and Piette, Bernard",
    title = "{Axially-symmetric sphaleron solutions of the Skyrme model}",
    doi = "10.1088/1742-6596/284/1/012053",
    journal = "J. Phys. Conf. Ser.",
    volume = "284",
    pages = "012053",
    year = "2011"
}

@article{Shnir:2013ova,
    author = "Shnir, Ya. and Zhilin, G.",
    title = "{Sphaleron solutions of the Skyrme model from Yang-Mills holonomy}",
    eprint = "1305.1456",
    archivePrefix = "arXiv",
    primaryClass = "hep-th",
    doi = "10.1016/j.physletb.2013.05.013",
    journal = "Phys. Lett. B",
    volume = "723",
    pages = "236--240",
    year = "2013"
}

@article{Klinkhamer:1984di,
    author = "Klinkhamer, Frans R. and Manton, N. S.",
    title = "{A Saddle Point Solution in the Weinberg-Salam Theory}",
    reportNumber = "NSF-ITP-84-57",
    doi = "10.1103/PhysRevD.30.2212",
    journal = "Phys. Rev. D",
    volume = "30",
    pages = "2212",
    year = "1984"
}

@article{Hamada:2020rnp,
    author = "Hamada, Yu and Kikuchi, Kengo",
    title = "{Obtaining the sphaleron field configurations with gradient flow}",
    eprint = "2003.02070",
    archivePrefix = "arXiv",
    primaryClass = "hep-th",
    reportNumber = "KUNS-2797, RIKEN-iTHEMS-Report-20",
    doi = "10.1103/PhysRevD.101.096014",
    journal = "Phys. Rev. D",
    volume = "101",
    number = "9",
    pages = "096014",
    year = "2020"
}

@article{Grant:2001at,
    author = "Grant, Jackie and Hindmarsh, Mark",
    title = "{Sphalerons in two Higgs doublet theories}",
    eprint = "hep-ph/0101120",
    archivePrefix = "arXiv",
    reportNumber = "SUSX-TH-01-003, SUSX-00-003",
    doi = "10.1103/PhysRevD.64.016002",
    journal = "Phys. Rev. D",
    volume = "64",
    pages = "016002",
    year = "2001"
}

@article{Adam:2021fet,
    author = "Adam, C. and Ciurla, D. and Oles, K. and Romanczukiewicz, T. and Wereszczynski, A.",
    title = "{Sphalerons and resonance phenomenon in kink-antikink collisions}",
    eprint = "2109.01834",
    archivePrefix = "arXiv",
    primaryClass = "hep-th",
    doi = "10.1103/PhysRevD.104.105022",
    journal = "Phys. Rev. D",
    volume = "104",
    number = "10",
    pages = "105022",
    year = "2021"
}

@article{Oles:2023ujf,
    author = "Oles, K. and Queiruga, J. and Romanczukiewicz, T. and Wereszczynski, A.",
    title = "{Sphaleron without shape mode and its oscillon}",
    eprint = "2309.11167",
    archivePrefix = "arXiv",
    primaryClass = "hep-th",
    doi = "10.1016/j.physletb.2023.138300",
    journal = "Phys. Lett. B",
    volume = "847",
    pages = "138300",
    year = "2023"
}

@article{Alonso-Izquierdo:2023shi,
    author = "Alonso-Izquierdo, A. and Navarro-Obreg{\'o}n, S. and Oles, K. and Queiruga, J. and Romanczukiewicz, T. and Wereszczynski, A.",
    title = "{Semi-Bogomol'nyi-Prasad-Sommerfield sphaleron and its dynamics}",
    eprint = "2308.14420",
    archivePrefix = "arXiv",
    primaryClass = "hep-th",
    doi = "10.1103/PhysRevE.108.064208",
    journal = "Phys. Rev. E",
    volume = "108",
    number = "6",
    pages = "064208",
    year = "2023"
}

@misc{Anco:2025egc,
    author = "Anco, Stephen C.",
    title = "{Instability of sphalerons in $\phi^4$ models with a false vacuum}",
    archivePrefix = {"arXiv"},
    eprint = {"2508.12150"},
    primaryClass = "hep-th",
    month = "8",
    year = "2025",
    note = {arXiv:2508.12150}
}

@article{Manton:2023afn,
    author = "Manton, N. S.",
    title = "{Integration theory for kinks and sphalerons in one dimension}",
    eprint = "2308.14453",
    archivePrefix = "arXiv",
    primaryClass = "hep-th",
    doi = "10.1088/1751-8121/ad14ac",
    journal = "J. Phys. A",
    volume = "57",
    number = "2",
    pages = "025202",
    year = "2024"
}

@article{Gleiser:1993pt,
    author = "Gleiser, Marcelo",
    title = "{Pseudostable bubbles}",
    eprint = "hep-ph/9308279",
    archivePrefix = "arXiv",
    reportNumber = "DART-HEP-93-05",
    doi = "10.1103/PhysRevD.49.2978",
    journal = "Phys. Rev. D",
    volume = "49",
    pages = "2978--2981",
    year = "1994"
}

@article{Gleiser:2009ys,
    author = "Gleiser, Marcelo and Sicilia, David",
    title = "{A General Theory of Oscillon Dynamics}",
    eprint = "0910.5922",
    archivePrefix = "arXiv",
    primaryClass = "hep-th",
    doi = "10.1103/PhysRevD.80.125037",
    journal = "Phys. Rev. D",
    volume = "80",
    pages = "125037",
    year = "2009"
}

@article{Lima:2021jxl,
    author = "Lima, Fred C. and Simas, Fabiano C. and Nobrega, K. Z. and Gomes, Adalto R.",
    title = "{Scattering of metastable lumps in a model with a false vacuum}",
    eprint = "2108.13579",
    archivePrefix = "arXiv",
    primaryClass = "hep-th",
    doi = "10.1016/j.physletb.2021.136707",
    journal = "Phys. Lett. B",
    volume = "822",
    pages = "136707",
    year = "2021"
}

@article{Dorey:2011yw,
    author = "Dorey, Patrick and Mersh, Kieran and Romanczukiewicz, Tomasz and Shnir, Yasha",
    title = "{Kink-antikink collisions in the $\phi^6$ model}",
    eprint = "1101.5951",
    archivePrefix = "arXiv",
    primaryClass = "hep-th",
    reportNumber = "DCPT-10-81",
    doi = "10.1103/PhysRevLett.107.091602",
    journal = "Phys. Rev. Lett.",
    volume = "107",
    pages = "091602",
    year = "2011"
}

@article{anco2026longtimebehavioursphaleronsphi4,
    author = "Stephen C. Anco and Danial Saadatmand",
    title = "{Long-time behaviour of sphalerons in $\phi^4$ models with a false vacuum}",
    eprint = "2510.04283",
    archivePrefix = "arXiv",
    primaryClass = "hep-th",
    doi = "10.1088/1751-8121/ae9a6f",
    journal = "J. Phys. A: Math. Theor.",
    volume = "59",
    pages = "345701",
    year = "2026"
}

@article{martinez2025oscillonsbubblesqballdynamics,
    author = "D. Canillas Martínez and P. Dorey and T. Romańczukiewicz and P. M. Saffin and K. Sławińska and A. Wereszczyński",
    title = "{Oscillons and bubbles in $Q$-ball dynamics}",
    eprint = "2509.03192",
    archivePrefix = "arXiv",
    primaryClass = "hep-th",
    doi = "10.1007/JHEP12(2025)154",
    journal = ". J. High Energ. Phys.",
    volume = "2025",
    pages = "154",
    year = "2025"
}

@misc{campos2026collisiondynamicsfalsevacuumoscillons,
      title={Collision Dynamics of False-Vacuum Oscillons}, 
      author={J. G. F. Campos and N. S. Manton and Azadeh Mohammadi},
      year={2026},
      eprint={2605.12633},
      archivePrefix={arXiv},
      primaryClass={hep-th},
      url={https://arxiv.org/abs/2605.12633},
      note = {arXiv:2605.12633}
}

@article{Amin:2010jq,
    author = "Amin, Mustafa A. and Shirokoff, David",
    title = "{Flat-top oscillons in an expanding universe}",
    eprint = "1002.3380",
    archivePrefix = "arXiv",
    primaryClass = "astro-ph.CO",
    doi = "10.1103/PhysRevD.81.085045",
    journal = "Phys. Rev. D",
    volume = "81",
    pages = "085045",
    year = "2010"
}

@article{Correa:2012gbo,
    author = "Correa, R. A. C. and de Souza Dutra, A.",
    title = "{Coupled Scalar Fields Oscillons and Breathers in Some Lorentz Violating Scenarios}",
    eprint = "1212.4448",
    archivePrefix = "arXiv",
    primaryClass = "hep-th",
    doi = "10.1155/2015/673716",
    journal = "Adv. High Energy Phys.",
    volume = "2015",
    pages = "673716",
    year = "2015"
}

\end{document}